\documentclass[11pt]{article}
\usepackage[utf8]{inputenc}
\usepackage{deauthor,times,graphicx}

\usepackage{multirow}
\usepackage{float}
\usepackage{caption}
\usepackage{amsmath}
\usepackage{graphicx, multicol, latexsym, amsmath, amssymb}
\usepackage{blindtext}
\usepackage{caption}

\PassOptionsToPackage{hyphens}{url}\usepackage{hyperref}

\newcommand{\cg}{ G}

\usepackage{booktabs,siunitx}

\newcommand{\pr}{{\tt \mathrm{Pr}}}
\newcommand{\hsys}{{\textsc{HypDB}}}
\newcommand{\sys}{{\textsc{Capuchin}}}

\newcommand{\mc}[1]{\mathcal{#1}}

\newcommand{\ignore}[1]{}

\newcommand*{\rom}[1]{\expandafter\@slowromancap\romannumeral #1@}

\newcommand{\indep}{\mbox{$\perp\!\!\!\perp$}}

\newcommand{\RNum}[1]{\uppercase\expandafter{\romannumeral #1\relax}}
\newcommand{\eg}{{\em e.g.}} 

\newcommand{\mb}[1]{{\mathbf{#1}}}

\newcommand{\proj}[1]{{\Pi}}
\newcommand{\sel}[1]{{\sigma}}

\newcommand{\cut}[1]{}
\newcommand{\eat}[1]{}

\begin{document}

\title{Data Management for Causal Algorithmic Fairness\footnote{This work is supported by the National Science Foundation under grants NSF III-1703281, NSF III-1614738, NSF AITF 1535565 and NSF award \#1740996.}}

\author{Babak Salimi\footnotemark[1], \  Bill Howe\footnotemark[2], \ Dan Suciu\footnotemark[1] \\
University of Washington\\
\footnotemark[1] {\{bsalimi,suciu\}@cs.washington.edu},  \footnotemark[2] {billhowe@uw.edu}}
\graphicspath{{salimi/}}

\maketitle

\begin{abstract}
Fairness is increasingly recognized as a critical component of machine learning systems.  However, it is the underlying data on which these systems are trained that often reflects discrimination, suggesting a data management problem.  In this paper, we first make a distinction between associational and causal definitions of fairness in the literature and argue that the concept of fairness requires causal reasoning. We then review existing works and identify future opportunities for applying data management techniques to causal algorithmic fairness.
\end{abstract}
\section{Introduction}
Fairness is increasingly recognized as a critical component of machine learning (ML) systems.  These systems are now routinely used to make decisions that affect people's lives~\cite{courtland2018bias}, with the aim of reducing costs, reducing errors, and improving objectivity. However, there is enormous potential for harm: The data on which we train algorithms reflects societal inequities and historical biases, and, as a consequence, the models trained on such data will therefore reinforce and legitimize discrimination and opacity.  The goal of  research on algorithmic fairness is to remove bias from machine learning algorithms.  

We recently argued that the algorithmic fairness problem is fundamentally a data management problem~\cite{salimi2019interventional}.  
The selection of sources, the transformations applied during pre-processing, and the assumptions made during training are all sensitive to bias that can exacerbate fairness effects. The goal of this paper is to discuss the application of data management techniques in algorithmic fairness. 
In Sec~\ref{sec:fairness} we make a distinction between associational and causal definitions of fairness in the literature and argue that the concept of fairness requires causal reasoning to capture natural situations, and that the popular associational definitions in ML can produce misleading results. In Sec~\ref{sec:dmandfairness} we review existing work and identify future opportunities for applying data management techniques to ensure causally fair ML algorithms.

\ignore{A naive (and ineffective) approach sometimes used in practice is to
simply omit the protected attribute (say, race or gender) when
training the classifier. However, since the protected attribute is
frequently represented implicitly by some combination of proxy variables, the
classifier still learns the discrimination reflected in training data.
For example, zip code tends to predict race
due to a history of segregation~\cite{amazonrace2016,selbst2017disparate}; answers to personality
tests identify people with disabilities~\cite{bodie2017law,personalitywsj2017}; and keywords can reveal gender on a resume~\cite{amazonhire2018}.
As a result, a classifier trained
without regard to the protected attribute not only fails to remove discrimination, but it can complicate the detection and mitigation of
discrimination downstream via in-processing or post-processing
techniques~\cite{russell2017worlds,galhotra2017fairness,corbett2017algorithmic,chouldechova2017fair,kusner2017counterfactual,kilbertus2017avoiding,nabi2018fair,Veale:2018:FAD:3173574.3174014}, which we next describe.}
\vspace{-0.3cm}
\section{Fairness Definitions}
\label{sec:fairness}
Algorithmic fairness considers a set of variables $\mb V$ that include a set of {\em protected attributes} $\mb S$ 
and a {\em response variable} $Y$, and a  classification algorithm
$\mc{A} : Dom(\mb X) \rightarrow Dom(O)$, where $\mb X \subseteq \mb V$,
and the result is denoted $O$
and called\ignore{ {\em output} or} {\em outcome}.  To simplify the exposition, we
assume a sensitive attribute $S \in \mb S$ that classifies the population into protected $S=1$ and
privileged $S=0$, for example, female and male, or minority and non-minority (see \cite{zliobaite2015survey} for a survey). 
The first task is to define formally when an algorithm $\mc{A}$ is fair w.r.t. the protected attribute $S$; such a definition is, as we shall see, not obvious.  Fairness definitions
can be classified as associational or causal, which we illustrate using the following running example (see \cite{verma2018fairness} for a survey on fairness definitions).

\begin{example} \em \label{ex:berkeley}
In 1973, UC Berkeley was sued for discrimination against
females in graduate school admissions. Admission figures for the fall of 1973 showed that men applying were more likely than women to be admitted, and
the difference was so large that it was unlikely to be due to chance.
However, it turned out that the observed correlation was due to the indirect effect of gender on admission results through applicant's choice of department. It was shown that 
	females tended to apply to departments with lower overall
	acceptance rates ~\cite{salimi2018bias}. 
	When broken down by department,  a slight bias toward female applicants
	was observed, a result that did not constitute evidence for gender-based discrimination. Extending this case, suppose college admissions decisions are made
	independently by each department and are based on a rich collection
	of information about the candidates, such as test scores, grades,
resumes, statement of purpose, etc.  These characteristics affect not
	only admission decisions, but also the department to which the
	candidate chooses to apply. The goal is to establish  conditions that guarantee fairness of admission decisions.
\end{example}

{\scriptsize
\begin{figure*}  \centering \scriptsize
	\begin{tabular}{|l|l|} \hline
		\textbf{     Fairness Metric}  & \textbf{Description}  \\ \hline
		Demographic Parity (DP) \cite{calders2009building}  & $S \indep O$  \\
		a.k.a. Statistical Parity \cite{dwork2012fairness} & \\
		or Benchmarking \cite{simoiu2017problem} & \\ \hline
		Conditional Statistical Parity \cite{corbett2017algorithmic} & $S \indep O |\mb A$ \\ \hline
		%
		Equalized Odds (EO) \cite{hardt2016equality} \footnotemark[2] & $S \indep O| Y $  \\
		a.k.a.  Disparate Mistreatment  \cite{zafar2017fairness} & \\ \hline
		Predictive Parity (PP)\cite{chouldechova2017fair} \footnotemark[3]  & $S \indep Y| O $  \\
		a.k.a. Outcome Test  \cite{simoiu2017problem} & \\
		or Test-fairness  \cite{chouldechova2017fair} & \\
		or Calibration \cite{chouldechova2017fair}, & \\
		or Matching Conditional Frequencies \cite{hardt2016equality} & \\ \hline
	\end{tabular}  
	\caption{      \textmd{ Common associational definitions of fairness.}} \label{tbl:asdef}
\end{figure*}
}
\vspace{-0.3cm}
\subsection{Associational Fairness} A simple and appealing approach to defining fairness is by correlating the sensitive attribute $S$ and the outcome $O$.  This leads to several possible definitions (Fig.~\ref{tbl:asdef}).
\textit{Demographic Parity} (DP)
\cite{dwork2012fairness} requires an algorithm to classify both
protected and privileged groups with the same probability, i.e.,
$\pr(O=1|S=1)=\pr(O=1|S=0)$.  However, doing so fails to correctly model our Example~\ref{ex:berkeley} since it requires equal probability for males and females to be admitted, and, as we saw, failure of DP cannot be considered evidence for gender-based discrimination.  This motivates  \textit{Conditional Statistical Parity} (CSP)
\cite{corbett2017algorithmic}, which controls for a set of
admissible factors  $\mb A$, i.e.,
$\pr(O=1|S=1,\mb A= \mb a)=\pr(O=1|S=0,\mb A=\mb a)$. 
The definition is  satisfied  if subjects in both
protected and privileged groups have equal probability of being
 assigned to the positive class, controlling for a set of admissible variables. In the UC Berkeley case, CSP is approximately satisfied by assuming that department is an admissible variable. 

Another popular measure used for predictive
classification algorithms is  \textit{Equalized Odds} (EO), which
requires both protected and privileged groups to have the same false
positive (FP) rate, $\pr(O=1|S=1,Y=0)=\pr(O=1|S=0,Y=0)$ , and the same
false negative (FN) rate, $\pr(O=0|S=1,Y=1)=\pr(O=0|S=0,Y=1)$, or,
equivalently, $(O \indep S|Y)$. In our example, assuming a classifier is trained to predict if an applicant will be admitted, then the false positive rate is the fraction of rejected applicants for which the classifier predicted that they should be admitted, and similarly for the false negative rate: EO requires that the rates of these false predictions be the same for male and female applicants. \ignore{The intuition behind PP is that the fraction of correct
positive predictions should be the same for both protected and privileged groups.} Finally, \textit{Predictive Parity} (PP)
requires that both protected and privileged groups have the same
predicted positive value (PPV),
$\pr(Y=1|O=i,S=0)=\pr(Y=1|O=i,S=1) \ \text{for} \ i,=\{1,0\}$ or,
equivalently, $Y \indep S| O$. In our example, this
implies that the probability of an applicant that actually got admitted  to be correctly classified as admitted
and the probability of an applicant that actually got rejected 
to be incorrectly classified as accepted should both be the same for male and female applicants.
\vspace{-0.3cm}
\paragraph{An Associational Debate.} Much of the literature in algorithmic fairness is motivated by controversies over a widely used commercial risk assessment system for recidivism --- COMPAS by Northpointe \cite{larson2016we}. In 2016, a team of journalists from ProPublica constructed a dataset of more than 7000 individuals arrested in Broward County, Florida between 2013 and 2014 in order to analyze the efficacy of COMPAS.  In addition, they collected data on arrests for these defendants through the end of March 2016. Their assessment suggested that COMPAS scores were biased against African-Americans based on the fact that the 
FP rate for African-Americans (44.9\%) was twice that for Caucasians (23.5\%). However, the FN rate for Caucasians (47.7\%) was twice as large as for  African-Americans (28.0\%). 
In other words, COMPAS scores were shown to violate EO.
In response to ProPublica, Northpointe showed
COMPAS scores satisfy PP, i.e., the likelihood of recidivism among high-risk offenders is the same regardless of race. 

This example illustrates that associational definitions are context-specific and can be mutually exclusive; they lack universality. Indeed, it has been shown that EO and PP are incompatible. In particular, Chouldechova~\cite{chouldechova2017fair} proves the following impossibility result.  Suppose that prevalence of the two populations differs,  $\pr(Y=1|S=0) \neq \pr(Y=1|S=1)$, for example, the true rate of recidivism differs for African-Americans and Caucasians; in this case, Equalized Odds and Predictive Parity cannot hold both simultaneously. Indeed, EO implies that $FP_i/(1-FN_i)$ is the same for both populations $S=i$, $i=0,1$, while PP implies that $(1-PPV_i)/PPV_i$ must be the same.  Then, the identity
{\small 
\begin{align*}
\frac{FP_i}{1-FN_i} = \frac{\pr(O=1|S=i,Y=0)}{\pr(O=1|S=i,Y=1)} =
 &  \frac{\pr(Y=1|S=i)}{\pr(Y=0|S=i)}  \frac{\pr(Y=0|O=1,S=i)}{\pr(Y=1|O=1,S=i)}
= \frac{\pr(Y=1|S=i)}{\pr(Y=0|S=i)} \frac{1-PPV_i}{PPV_i}
\end{align*}
}
for $i=0,1$, implies $\pr(Y=1|S=0) = \pr(Y=1|S=1)$.
\ignore{
when 
prevalence of the two
populations differs, meaning $\pr(Y=1|S=0) \neq \pr(Y=1|S=1)$.  The
proof follows immediately from her observation that, for each
population group $S=i$, the following holds: \footnote{EO implies
$FP/(1-FN)$ is the same for both groups,
$\frac{\pr(O=1|S=0,Y=0)}{\pr(O=1|S=0,Y=1)}=\frac{\pr(O=1|S=1,Y=0)}{\pr(O=1|S=1,Y=1)}$,
while PP implies that $(1-PPV)/PPV$ is the same for both groups,
$\frac{\pr(Y=0|O=1,S=0)}{\pr(Y=1|O=1,S=0)}=\frac{\pr(Y=0|O=1,S=1)}{\pr(Y=1|O=1,S=1)}$.
When the prevalence differs, EO and PP cannot hold simultaneously.}
\begin{align}
 \frac{\pr(O=1|S=i,Y=0)}{\pr(O=1|S=i,Y=1)} =
 &  \frac{\pr(Y=1|S=i)}{\pr(Y=0|S=i)}  \frac{\pr(Y=0|O=1,S=i)}{\pr(Y=1|O=1,S=i)}
\end{align}
}
We revisit the impossibility result in Sec~\ref{sec:imposib}.
\vspace{-0.3cm}

\subsection{Causal Fairness}

\label{sec:cf}
The lack of universality and the impossibility result for fairness definitions based on associational definitions have motivated definitions based on causality~\cite{kusner2017counterfactual,kilbertus2017avoiding,nabi2018fair,russell2017worlds,galhotra2017fairness}.  The intuition is simple: fairness holds when there is no causal relationship from the protected attribute $S$ to the outcome $O$.  We start with a short background on causality.

\ignore{
was motivated by the need to address the issues with
associational fairness and assumes an underlying causal model.  We
first discuss causal models before reviewing causal fairness.
\vspace{-0.3cm}
\subsection{Background on Causal DAGs }

\label{sec:causal:dag}

}

\vspace{-0.3cm}
\paragraph*{\bf Causal DAG.}
A \textit{causal DAG} $\cg$ over a set of variables $\mb V$ is a directed acyclic graph that
models the functional interaction between variables in $\mb V$.  Each node $X$ represents
a variable in $\mb V$ that is functionally determined by: (1) its parents $\mb{Pa}(X)$
in the DAG, and (2) some set of \emph{exogenous} factors that need not appear in the DAG
as long as they are mutually independent.
This functional interpretation leads to the same decomposition of the joint probability distribution of $\mb V$ that characterizes Bayesian networks \cite{pearl2009causality}:
\begin{align}
\pr(\mb V) = & \prod_{X \in \mb V} \pr(X | \mb{Pa}(X)) \label{eq:bayesian1}
\end{align}
\vspace{-0.3cm}
\paragraph*{\bf $d$-Separation.}
A common inference question in a causal DAG is how to determine whether a
CI $(\mb X \indep \mb Y | \mb Z)$ holds.  A sufficient criterion is
given by the notion of d-separation,  a syntactic condition
$(\mb X \indep \mb Y |_d \mb Z)$ that can be checked directly on the
graph (we refer the reader to \cite{pearl2003causality} for details).  \ignore{$\pr$ and $\cg$ are
called {\em Markov compatible} if $(\mb X \indep \mb Y |_d \mb Z)$
implies $(\mb X \indep_\pr \mb Y | \mb Z)$; if the converse
implication holds, then we say that $\pr$ is {\em faithful} to $\cg$.
It is known that if $\cg$ is a causal DAG and
  $\Pr$ is given by Eq.(\ref{eq:bayesian1}), then they are Markov
  compatible}

\ignore{
\begin{figure*} 
	\hspace*{.8cm}	\begin{minipage}{0.4\textwidth} 
	\hspace*{.8cm}	\includegraphics[scale=0.15]{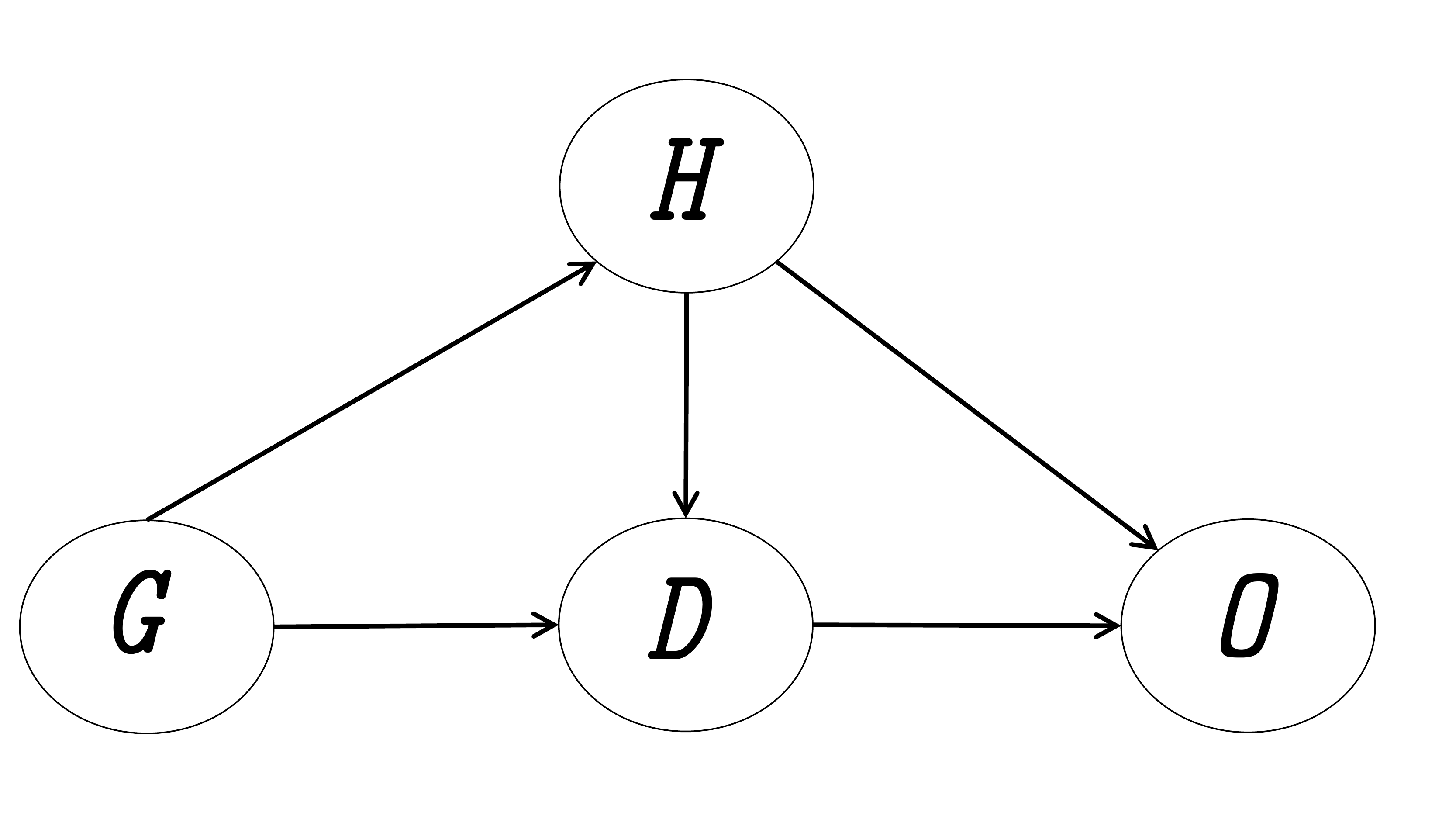}
	\caption*{(a)}
	\end{minipage}
	\begin{minipage}{0.4\textwidth} 
	\hspace*{.8cm}	\includegraphics[scale=0.15]{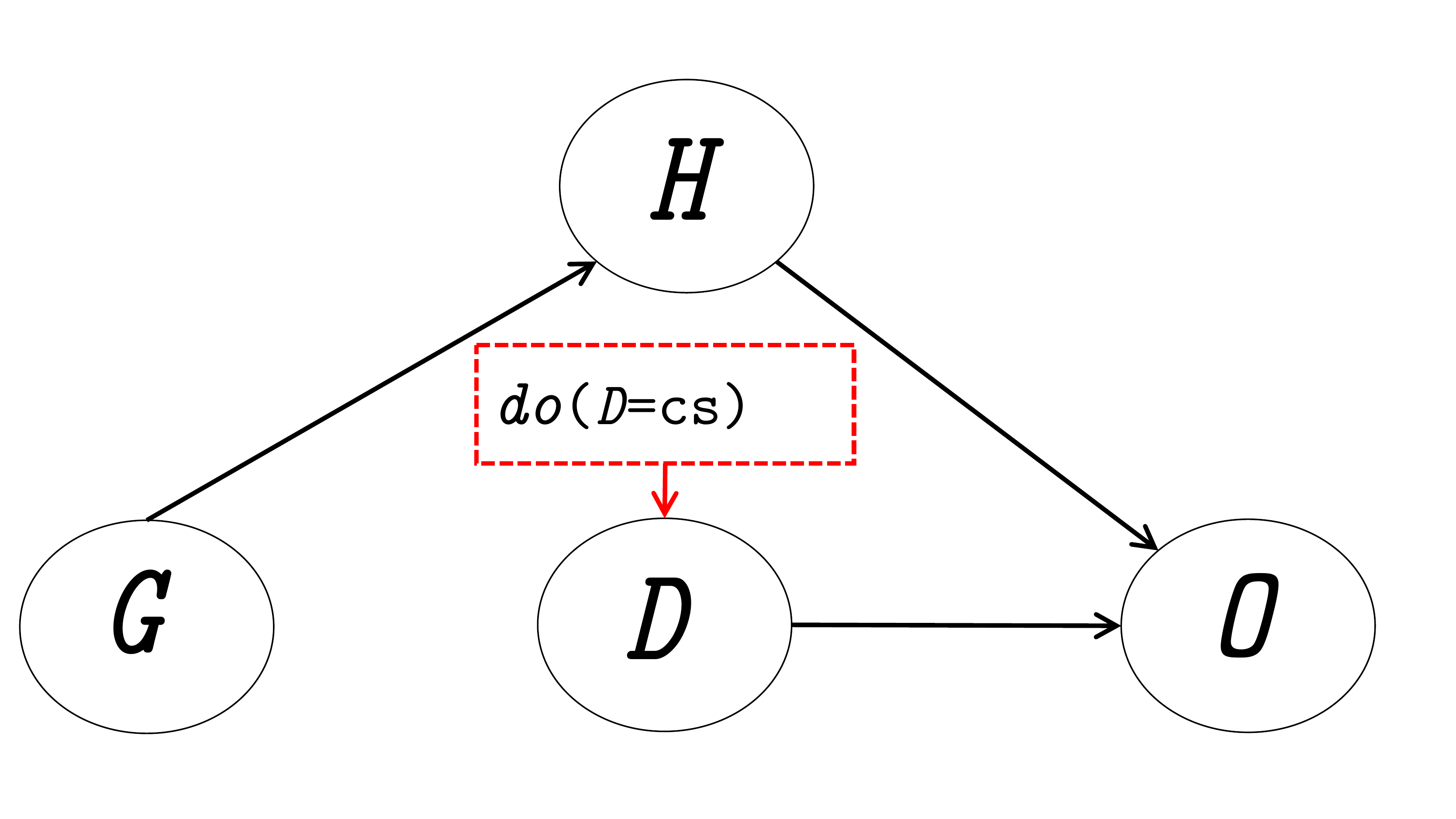}
	\caption*{(b)}
	\end{minipage}
	\caption{(a) Represents a causal DAG with $A=$ admission outcome, $G=$ applicant's gender,
		$H=$ applicant's hobbies, and $D=$ applicant's choice of department (cf. Ex.~\ref{ex:berkeley}).
		(b) Represents  the causal DAG obtained after external interventions (cf. Ex.~\ref{ex:berkeley}).}
	\label{fig:cgex}
\end{figure*}
}
\vspace{-0.3cm}
\paragraph*{\bf Counterfactuals and \texttt{do} Operator.}
A \textit{counterfactual} is an intervention where we actively modify
the state of a set of variables $\mb X$ in the real world to some
value $\mb X= \mb x$ and observe the effect on some output $Y$.
Pearl~\cite{pearl2009causality} described the $do$ operator, which
allows this effect to be computed on a causal DAG, denoted
$\pr(Y|do(\mb X= \mb x))$.  To compute this value, we assume that $X$ is
determined by a constant function $\mb X= \mb x$ instead of a function provided
by the causal DAG.  This assumption corresponds to a modified graph
with all edges into $\mb X$ removed, and values of the incoming variables are
set to $\mb x$.  For a simple example, consider three random variables $X,Y,Z\in \{0,1\}$.  We randomly flip a coin and set $Z=0$ or $Z=1$ with probability $1/2$; next, we set $X=Z$, and finally we set $Y=X$.  The resulting causal DAG is $Z\rightarrow X \rightarrow Y$, whose equation is $\pr(X,Y,Z) = \pr(Z)\pr(X|Z)\pr(Y|X)$.  The $do$ operator lets us observe what happens in the system when we intervene by setting $X=0$. The result is defined by removing the edge $Z \rightarrow X$, whose equation is $\pr(Y=y,Z=z|do(X)=0) = \pr(Z=z)\pr(Y=y|X=0)$ (notice that $\pr(X|Z)$ is missing), leading to  the marginals $\pr(Y=0|do(X)=0)=1, \pr(Y=1|do(X)=0)=0$.  It is important to know the casual DAG since the probability distribution is insufficient to compute the $do$ operator; for example, if we reverse the arrows to $Y\rightarrow X \rightarrow Z$ (flip $Y$ first, then set $X=Y$, then set $Z=X$), then $\pr(Y=0|do(X)=0)=\pr(Y=1|do(X)=0)=1/2$ in other words, intervening on $X$ has no effect on $Y$.

\ignore{The Bayesian rule Eq.(\ref{eq:bayesian1}) for the
modified graph defines $\pr(Y|do(\mb X=\mb x))$ and is given by the following extended adjusted formula \cite{salimi2019interventional}:
\begin{theorem}  \label{theo:af} Given a causal DAG $\cg$ and a set of
  variables $\mb X \subseteq \mb V$, suppose
  $\mb X=\{X_0,X_1 \ldots X_m \}$ are ordered such that $X_i$ is a
  non-descendant of $X_{i+1}$ in $\cg$. The effect of a set of
  interventions $do(\mb X= \mb x )$ is given by the following {\em
    extended adjustment formula}:
{\small
	\begin{align}
\pr(y|do(\mb X= \mb x )) &=& \nonumber \\
    &&  \hspace*{-2cm}\sum_{ \mb z \in Dom(\mb Z)} \pr( y | \mb x, \mb z )  \bigg( \prod_{i=0}^{m} \pr\big(\mb{pa}(X_i)\bigg|  \bigcup_{j=0}^{i-1} \mb{pa}(X_j ), \bigcup_{j=0}^{i-1}x_j  \big) \bigg) \label{eq:af}
	\end{align}
}
	\noindent where $\mb Z= \bigcup_{X \in \mb X} \mb{Pa}(X)$ and
	$j\geq 0$.
\end{theorem}
}
\vspace{-0.2cm}
  \paragraph*{\bf Counterfactual Fairness.}
 Given a set of features $\mb X$, a protected attribute $S$, an outcome variable $Y$, and a set of unobserved exogenous  background  variables $\mb U$, Kusner et
	al.~\cite{kusner2017counterfactual} defined a predictor $O$
	to be {\em counterfactually fair} if for any $\mb x \in Dom(\mb X)$:
	%
	{ 
		\begin{align}
		P( O_{S\leftarrow 0}(\mb U) = 1| \mb X = \mb x,S = 1) = P( O_{S\leftarrow 1}(\mb U)= 1| \mb X = \mb x;S = 1) \label{eq:cfair}
		\end{align}
	}
	\noindent  where $O_{S\leftarrow s}(\mb U)$ means intervening on
	the protected attribute in an unspecified configuration of the exogenous factors.
	The definition is meant to capture the requirement that the protected attribute $S$ should not be a cause of $O$ at the individual level. However, this definition captures individual-level fairness
  only under certain strong assumptions (see \cite{salimi2019interventional}).
  Indeed, it is known in statistics that individual-level
  counterfactuals cannot be estimated from data \cite{rubin1970thesis,rubin1986statistics,rubin2008comment}.  

\vspace{-0.3cm}
  \paragraph*{\bf Proxy Fairness.} To avoid individual-level
  counterfactuals, a common approach is to study
  population-level counterfactuals or interventional distributions
  that capture the effect of interventions at population rather
  than individual level
  \cite{pearl2009causal,rubin1970thesis,rubin1986statistics}.
  Kilbertus et al. \cite{kilbertus2017avoiding} defined proxy
  fairness as follows:
  {
    \begin{align}
      P(O=1| do(\mb P=\mb p ))=P(O=1| do(\mb  P=\mb p')) \label{eq:pfair}
    \end{align}
  }
  \noindent for any $\mb p, \mb p' \in Dom(\mb P)$, where $\mb P$
  consists of proxies to a sensitive variable $S$ (and might include
  $S$).  Intuitively, a classifier satisfies proxy fairness in
  Eq~\ref{eq:pfair} if the distribution of $O$ under two
  interventional regimes 
  in which $\mb P$ set to $\mb p$ and $\mb p'$
  is the same. Thus, proxy fairness is not an individual-level notion.
It has been shown that proxy fairness fails to capture group-level
  discrimination in general \cite{salimi2019interventional}.  

\paragraph*{\bf Path-Specific Fairness.} These definitions are based on
graph properties of the causal graph, \eg, prohibiting specific paths from the
sensitive attribute to the outcome~\cite{nabi2018fair,loftus2018causal}; however,
identifying path-specific causality from data requires very strong assumptions and is often impractical~\cite{avin2005identifiability}.
\vspace{-0.3cm}
\paragraph*{\bf Interventional Fairness.} To avoid issues with the aforementioned causal definitions, Salimi et al. \cite{salimi2019interventional} defined interventional fairness as follows: an
  algorithm $\mc A : Dom(\mb X) \rightarrow Dom(O)$ is $\mb K$-fair for a set of
  attributes $\mb K \subseteq \mb V-\{S,O\}$  w.r.t. a protected attribute $S$ if, for any context $\mb K = \mb k$
  and every outcome $O=o$, the following holds:
	{ 
	\begin{eqnarray}
	\pr(O=o| do(S=0),do(\mb K = \mb k))= \pr(O=o| do(S=1), do(\mb K = \mb k)) \label{eq:cfair}
	\end{eqnarray}
}
\noindent An algorithm is called {\em interventionally fair} if it is
  $\mb K$-fair for every set $\mb K$.  Unlike proxy fairness, this
  notion correctly captures group-level fairness because it ensures
  that $S$ does not affect $O$ in \emph{any configuration} of the
  system obtained by fixing other variables at some arbitrary values.
  Unlike counterfactual fairness, it does not attempt to capture
  fairness at the individual level, and therefore it uses the standard
  definition of intervention (the \texttt{do}-operator).  In practice, interventional fairness is too restrictive. For example, in the UC Berkeley case, admission decisions were not interventionally fair since gender affected the admission result via applicant's choice of department. To make it practical, Salimi et al. \cite{salimi2019interventional} defined a notion of fairness that relies on partitioning variables into {\em admissible} and {\em inadmissible}.  The former are variables
through which it is permissible for the protected attribute to
influence the outcome.  This partitioning expresses fairness social norms and values and comes from the users. In Example~\ref{ex:berkeley}, the user would
label department as admissible since it is considered a fair use in
admissions decisions and would (implicitly) label all other variables
as inadmissible, for example, hobby. Then, an algorithm is called {\em justifiably fair} if it is $\mb K$-fair w.r.t. all supersets $\mb K \supseteq \mb A$.  We illustrate with an example.
\begin{figure*} \centering
	\begin{minipage}{0.3\textwidth} 
		\includegraphics[scale=0.15]{Fig2_a.pdf}
		\caption*{(a) College~\RNum{1}}
	\end{minipage}
	\begin{minipage}{0.3\textwidth} 
		\includegraphics[scale=0.15]{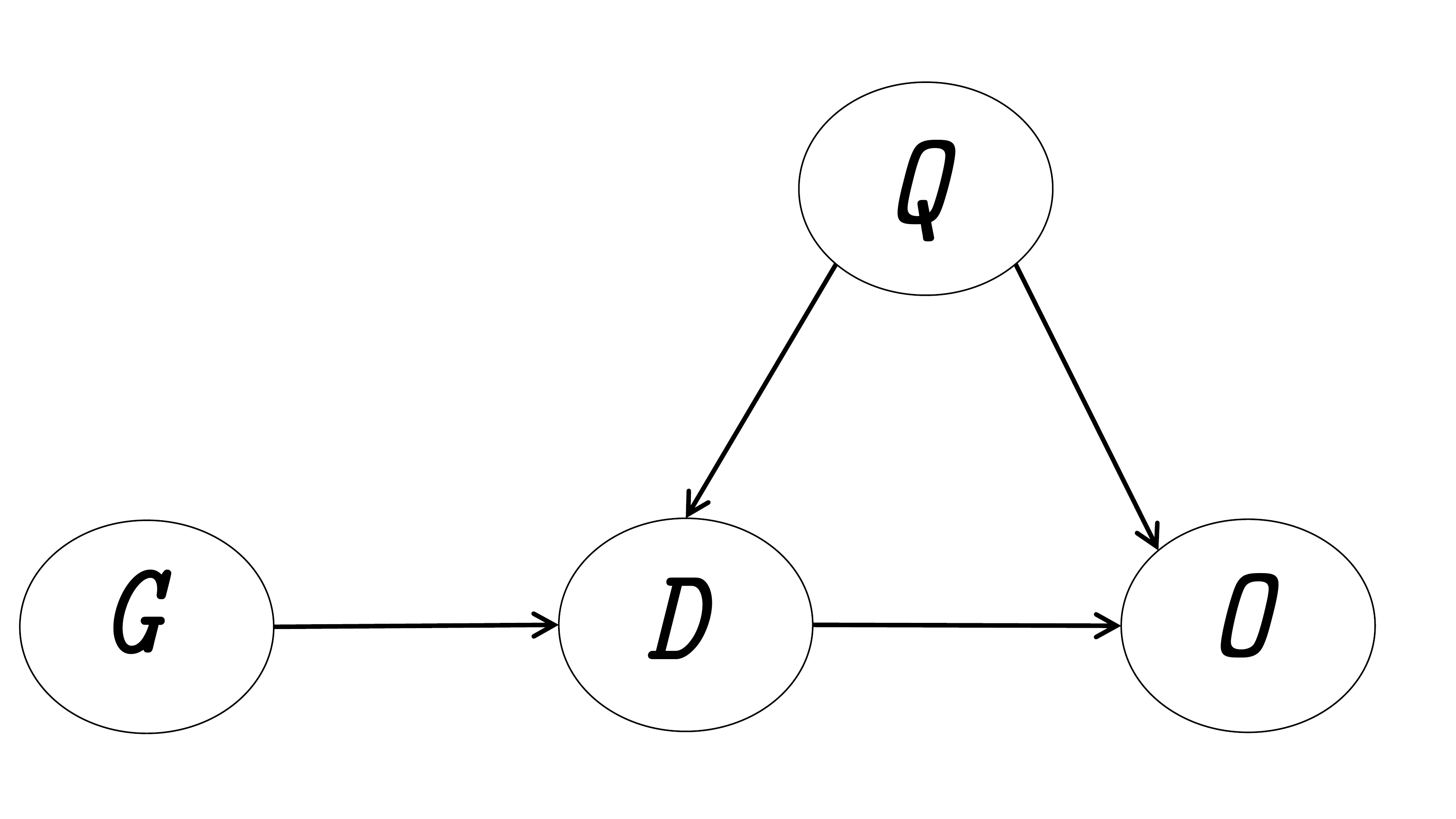}
		\caption*{(b) College~\RNum{2}}
	\end{minipage}

	{ \small
		\begin{tabular}{c@{\qquad}ccc@{\qquad}ccc}
			\toprule
			\multirow{2}{*}{\raisebox{-\heavyrulewidth}{{\bf College~\RNum{1}}}} & \multicolumn{2}{c}{Dept.~A} & \multicolumn{2}{c}{Dept.~B} & \multicolumn{2}{c}{Total}  \\
			\cmidrule{2-7}
			& Admitted & Applied & Admitted &Applied & Admitted & Applied \\
			\midrule
			Male& 16 &  20 & 16 & 80 & 32 & 100 \\
			Female& 16 & 80 & 16 & 20 & 32 & 100 \\
			\bottomrule
		\end{tabular}
\hspace*{-0.2cm}			\begin{tabular}{c@{\qquad}ccc@{\qquad}ccc}
			\toprule
			\multirow{2}{*}{\raisebox{-\heavyrulewidth}{\bf College~~\RNum{2}}} & \multicolumn{2}{c}{Dept.~A} & \multicolumn{2}{c}{Dept.~B} & \multicolumn{2}{c}{Total}  \\
			\cmidrule{2-7}
			& Admitted & Applied & Admitted &Applied & Admitted & Applied \\
			\midrule
						Male& 10 & 10 & 40 & 90 & 50 & 100 \\
			Female& 40 &  50 & 10 & 50 & 50 & 100 \\

			\bottomrule
		\end{tabular}
	}
  
	\caption{      \textmd{ Admission process representation in two colleges where  associational fairness fail (see Ex.\ref{ex:ap_str}).}}
	\label{adm_rate}
\end{figure*}
\vspace{-0.2cm}
\begin{example} \label{ex:ap_str}  Fig~\ref{adm_rate} shows how fair or
  unfair situations may be hidden by coincidences but exposed through
  causal analysis.  In both examples, the protected attribute is
  gender $G$, and the admissible attribute is department $D$. Suppose
  both departments in College~\RNum{1} are admitting only on the basis
  of  their applicants' hobbies.  Clearly, the admission process is
  discriminatory in this college because department~A admits 80\% of
  its male applicants and 20\% of the female applicants, while
  department~B admits 20\% of male and 80\% of female applicants.  On
  the other hand, the admission rate for the entire college is the
  same 32\% for both male and female applicants, falsely suggesting
  that the college is fair.  Suppose $H$ is a proxy to $G$ such that $H=G$ ($G$ and $H$ are the same); proxy
  fairness then classifies this example as
  fair: indeed, since Gender has no parents in the causal graph,
  intervention is the same as conditioning; hence,
  $\Pr(O=1|do(G=\text{i}))=\pr(O=1| G = \text{i})$ for $i=0,1$.  Of the previous methods, only conditional
  statistical parity correctly indicates discrimination.  We
  illustrate how our definition correctly classifies this examples as
  unfair.  Indeed, assuming the user labels the department $D$ as
  admissible, $\{D\}$-fairness fails because
  $\pr(O=1|do(G=1),do(D=\text{'A'}))=\sum_h
  \pr(O=1|G=1,D=\text{'A'},H=h)\pr(H=h|G=1) =
  \pr(O=1|G=1,D=\text{'A'})=0.8$,
%
%
%
%
  and, similarly $\pr(O=1|do(G=0),do(D=\text{'A'}))=0.2$.
  Therefore, the admission process is not justifiably fair.

  Now, consider the second table for College~\RNum{2}, where both
  departments~A and B admit only on the basis of student
  qualifications $Q$.  A superficial examination of the data suggests
  that the admission is unfair: department A admits 80\% of all
  females and 100\% of all male applicants; department B admits 20\%
  and 44.4\%, respectively.  Upon deeper examination of the causal DAG,
  we can see that the admission process is justifiably fair because
  the only path from Gender to Outcome goes through Department,
  which is an admissible attribute.  To understand how the data could
  have resulted from this causal graph, suppose 50\% of each gender
  have high qualifications and are admitted, while others are
  rejected, and that 50\% of females apply to each department, but
  more qualified females apply to department A than to B (80\%
  vs 20\%). Further, suppose fewer males apply to department A, but
  all of them are qualified.  The algorithm satisfies demographic
  parity and proxy fairness but fails to satisfy conditional
  statistical parity since $\pr(A=1|G=1,D=\text{A})=0.8$ but
  $\pr(A=1|G=0,D=\text{A})=0.2)$.  Thus, conditioning on $D$ falsely
  indicates discrimination in College~\RNum{2}. One can check that the
  algorithm is justifiably fair, and thus our definition also
  correctly classifies this example; for example, $\{D\}$-fairness
  follows from
  $\pr(O=1|do({G}={i}),do(D=d))= \sum_{q}\pr(O=1|G=i,D=d,Q=q))
  \nonumber \pr(Q=q|G=i)= \frac{1}{2}$.
 To summarize, unlike previous definitions of fairness, justifiable  fairness correctly identifies College~\RNum{1} as discriminatory and College~\RNum{2} as fair.

\end{example}

\subsection{Impossibility Theorem from the Causality Perspective}
\label{sec:imposib}
From the point of view of causal DAGs, EO requires that the training label $Y$
$d$-separates the sensitive attribute $S$ and the outcome of the classifier $O$. Intuitively, this implies that $S$ can affect classification results only when the information comes through the training label $Y$. On the other hand, PP requires that the classifier outcome $O$ $d$-separates the sensitive attribute $S$ and the training labels $Y$. Intuitively, this implies $S$ can affect the training labels only when the information comes thorough the outcome of classifier $O$. These interpretations clearly reveal the inconsistent nature of EO and PP. It is easy to show for strictly positive distributions that the CIs ($S \indep O| Y$) and  ($S \indep Y| O$) imply ($S \indep Y$) 
or, equivalently,  $\pr(Y=1|S=0) = \pr(Y=1|S=1)$ (see \cite{salimi2019interventional}).  Indeed, from the causality perspective, EO and PP are neither sufficient nor necessary for fairness. In the causal DAG in Fig~\ref{fig:cgex}(b), suppose a classifier is trained on an applicant's qualifications $Q$ to approximate admission committee decisions $\hat{O}$.  It is clear that the classifier is not discriminative, yet it violates both EO and PP. The reader can verify that the causal DAG obtained by further adding an edge from $Q$ to $\hat{O}$ (to account for the classifier outcome) does not imply the CIs ($G \indep O| \hat{O}$) and ($G \indep \hat{O}| O)$.
\vspace{-0.3cm}
\section{Data Management Techniques for Causal Fairness}
\label{sec:dmandfairness}
\vspace{-0.1cm}
\subsection{Causal Fairness as Integrity Constraints}
\label{sec:integ}
In causal DAGs, the missing arrow between two variables $X$ and $Y$ represents the assumption of no causal effect between them, which corresponds to the CI statement ($X \indep Y | \mb Z$), where $\mb Z$ is a set of variables that $d$-separates $X$ and $Y$. For example, the missing arrow between $O$ and $G$  in the causal DAG in Fig.~\ref{adm_rate}(a) encodes the CI $(O \indep G| H,D)$. On the other hand, the lack of certain arrows in the underling causal DAG is sufficient to satisfy different causal notions of fairness (cf. Sec~\ref{sec:cf}). For instance, a sufficient condition for justifiable fairness in the causal DAG in  Fig.~\ref{adm_rate}(a) is the lack of the edge from $H$ to $O$, which corresponds to the CI $(O \indep G, H| D)$. Thus, fairness can be captured as a set of CI statements. Now to enforce fairness, instead of intervening on the causal DAG over which we have no control, we can intervene on data to enforce the corresponding CI statements. 

Consequently, social causal fairness constraints can be seen as a set of integrity constraints in the form of CIs that must be preserved and enforced thorough the data science pipeline, from data gathering through the deployment of a machine learning model. The connection between CIs and well-studied integrity constraints in data management -- such as Multi Valued Dependencies (MVDs) and Embedded Multi Valued Dependencies (EMVDs)~\cite{AbiteboulHVBook} -- opens the opportunity to leverage existing work in data management to detect and avoid bias in data.


\vspace{-0.3cm}
\subsection{Query Rewriting}
In data management, \textit{query rewriting} refers to a set of techniques to automatically modify one query into another that satisfies certain desired properties. These techniques are used to rewrite queries with views~\cite{DBLP:conf/pods/LevyMSS95}, in chase and backchase for complex optimizations~\cite{DBLP:conf/sigmod/PopaDST00}, and for many other applications.  This section discusses query rewriting techniques for detecting and enforcing fairness.

\begin{figure*} 
	\includegraphics[scale=0.51]{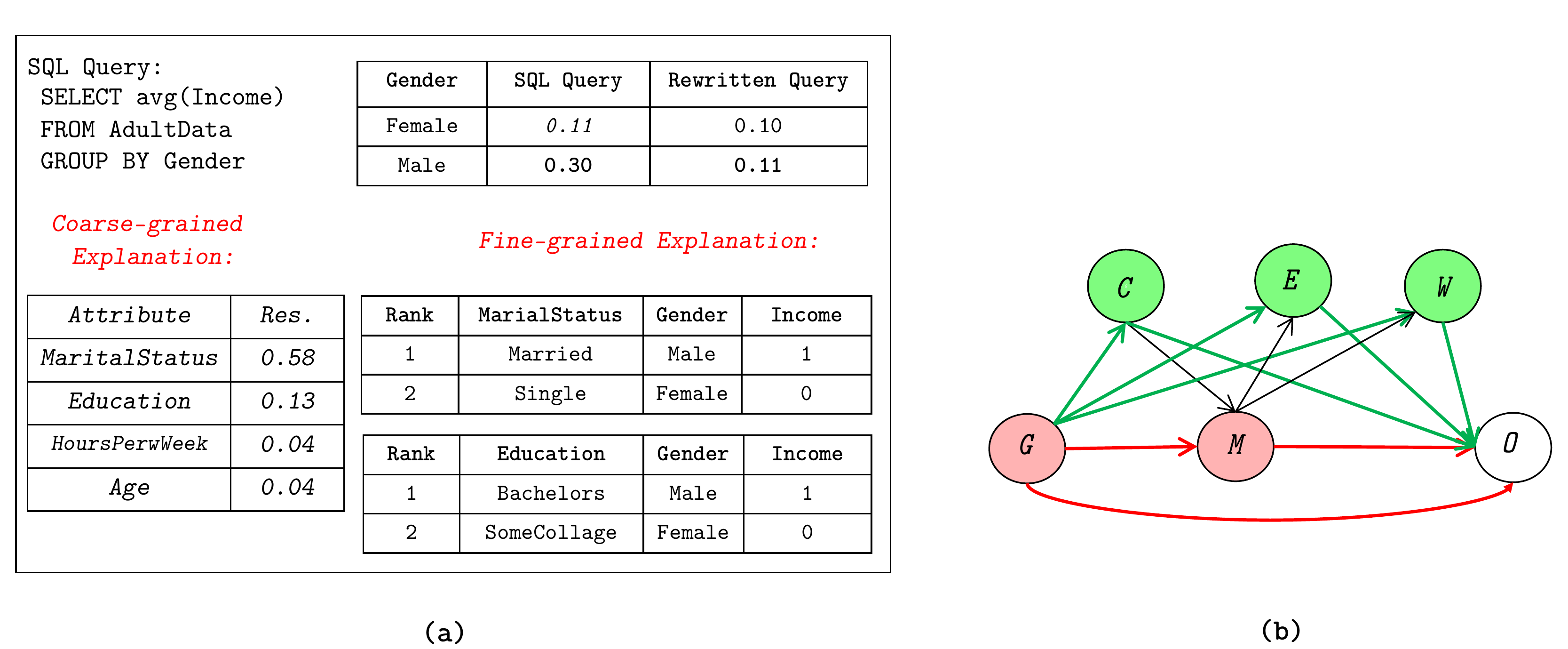}
	\caption{(a) \hsys's report on the effect of gender on income (cf.~Ex.~\ref{ex:berkeley}).
		(b) A compact causal DAG with $O=$ income, $G=$  gender,
		$M=$ marital status, $C=$ age and nationality, $E=$ education and $W=$ work class, occupation and hours per week (cf. Ex.~\ref{ex:adult}).}
	\label{fig:cgex}
\end{figure*}
\vspace{-0.3cm}
\subsubsection{Detecting Discrimination}
\label{sec:detct}
As argued in Sec~\ref{sec:cf}, detecting discrimination should rely on performing a hypothesis
test on the causal effect of membership in minority $S=1$ or 
privileged group $S=0$ on an outcome of an algorithm $O$.
The gold standard for such causal hypothesis testing is a {\em
    randomized experiment} (or an {\em A/B test}), called such
  because treatments are randomly assigned to subjects.  In contrast, in the context of fairness, sensitive attributes are typically imputable; hence, randomization is not even conceivable. Therefore, such queries must be answered using {\em observational data}, defined as data recorded from the environment with no randomization or other controls.  Although causal inference in observational data has been studied in
statistics for decades, causal analysis is not supported in existing online analytical processing (OLAP) tools \cite{salimi2018bias}. Indeed, today, most data analysts still reach for the simplest query that computes the average of $O$ Group By $S$ to answer such questions, which, as shown in Ex~\ref{ex:berkeley}, can lead to incorrect conclusions. Salimi et al. \cite{salimi2018bias} took the first step toward extending existing OLAP tools to support causal analysis. Specifically, they introduced the \hsys\ system, which brings together techniques from data management and causal inference to automatically rewrite SQL group-by queries into complex causal queries that support decision making. We illustrate \hsys\ by applying it to a fairness question (see \cite{salimi2018hypdb} for additional examples). 
\begin{example} \label{ex:adult}  Using UCI adult Census data \cite{adult}, several prior works in algorithmic fairness have reported gender discrimination based on the fact that 11\% of women have
	high income compared to 30\% of men, which suggests a huge disparity against women.  To decide whether the observed strong correlation between gender and high income is due to discrimination, we need to understand its causes.
	To perform this analysis using \hsys, one can start with the simple group-by query (Fig. \ref{fig:cgex}(a)) that computes the average of Income (1 iff Income$>$ 50k) Group By Gender, which indeed suggests a strong
disparity with respect to females' income. While the group-by query tells us gender and high income are highly correlated, it does not tell us why. To answer this question, \hsys\ automatically infers from data that gender can potentially influence income indirectly via  MaritalStatus,  Education, Occupation, etc. (the indirect causal paths from $G$ to $O$ in Fig.~\ref{fig:cgex}(b)). Then, \hsys\ automatically
rewrites the group-by query to quantify the direct and indirect effect of gender on income. Answers to the rewritten queries suggest that the direct effect of gender on income is not significant (the effect through the arrow from $G$ to $O$ in Fig. \ref{fig:cgex}(b)). Hence, gender essentially influences income indirectly through mediating variables. To understand the nature of this influences, \hsys\ provides the user with several explanations. These show that MaritalStatus accounts for most of the indirect influence, followed by Education.  However, the top fine-grained explanations for MaritalStatus reveal  surprising facts: there are more married males in the data than married females, and marriage has a strong positive association with high income. It turns out that the income attribute in US census data reports the adjusted gross income as indicated in the individual's tax forms; 
these depend on  filing status (jointly and separately), could be  household income. \hsys\  explanations also show that males tend  to have higher levels of education than females, and higher levels of education is associated with higher incomes. The explanations generated by \hsys\ illuminate crucial factors for investigating gender discrimination.
\end{example}
\vspace{-0.5cm}
\paragraph{Future Extensions.} Incorporating the type of analyses supported by \hsys\ into data-driven decision support systems is not only crucial for sound decision making in general, but it is also important for detecting, explaining and avoiding bias and discrimination in data and analytics. Further research is required on extending \hsys\ to support more complex types of queries and data, such as multi-relational and unstructured.
\vspace{-0.3cm}
\subsubsection{Enforcing Fairness}
\label{sec:queryrep}

Raw data often goes through a series of transformations to enhance the clarity and relevance of the signal used for a particular machine learning application \cite{antenucci2018constraint}.  Filter transformations are perhaps most common, in which a subset of training data is removed based on predicates. Even if the raw data is unbiased, filtering can introduce bias \cite{antenucci2018constraint,salimi2018bias}: It is known that causal DAGs are not closed under conditioning because CIs may not hold in some subset. Hence, filtering transformations can lead to violation of causal fairness integrity constraints. It is also known that conditioning on common effects can further introduce bias even when the sensitive attribute and training labels are marginally independent \cite{pearl2003causality}.  This motivates the study of {\em fairness-aware data transformations}, where the idea is to minimally rewrite the transformation query so certain fairness constraints are guaranteed to be satisfied in the result of the transformation. This problem is closely related to that of constraint-based data transformations studied in \cite{antenucci2018constraint}. However, fairness constraints go beyond the types of constraints considered in \cite{antenucci2018constraint} and are more challenging to address. Note that a solution to the aforementioned problem can be used to enforce fairness-constraints for raw data by applying a fair-transformation that selects all the data. 


\vspace{-0.3cm}
\subsection{Database Repair}
\label{sec:dbrepair}

Given a set of integrity constraints $\Gamma$ and a database instance $D$ that is inconsistent with $\Gamma$, the problem of repairing $D$ is to find an instance $D'$ that is close to $D$ and consistent with $\Gamma$. Repair of a database can be obtained by deletions
and insertions of whole tuples as well as by updating attributes. The closeness between $D$ and $D'$ can be interpreted in many
different ways, such as the minimal number of changes or the minimal set of changes under set inclusion (refer to \cite{DBLP:series/synthesis/2011Bertossi} for a survey). The problem has been studied extensively in database theory for
various classes of constraints.  It is NP-hard even
when $D$ consists of a single relation
and $\Gamma$ consists of functional
dependencies~\cite{DBLP:conf/pods/LivshitsKR18}.  

Given a training data $D$ that consists of a training label $Y$, a set of admissible variables $\mb{A}$, and a set of inadmissible variables $\mb{I}$, Salimi et al \cite{salimi2019interventional} showed that a sufficient condition for a classifier to be justifiably fair is that the empirical distribution $\pr$ over $D$ satisfies the CI $(Y \indep \mb I |\mb{A})$.  Further, they introduced the \sys\ system, which minimally repairs $D$ by performing a sequence of database updates (viz., insertions and deletions of
tuples) to obtain another training database $D'$ that satisfies 
$(Y \indep \mb I |\mb{A})$. Specifically, they reduced the problem to a minimal repair problem w.r.t. an MVD and developed a set of techniques, including reduction to the MaxSAT and Matrix Factorization, to address the corresponding optimization problem. We illustrate \sys\ with an example.

\begin{figure}[]  \centering
	\includegraphics[scale=0.5]{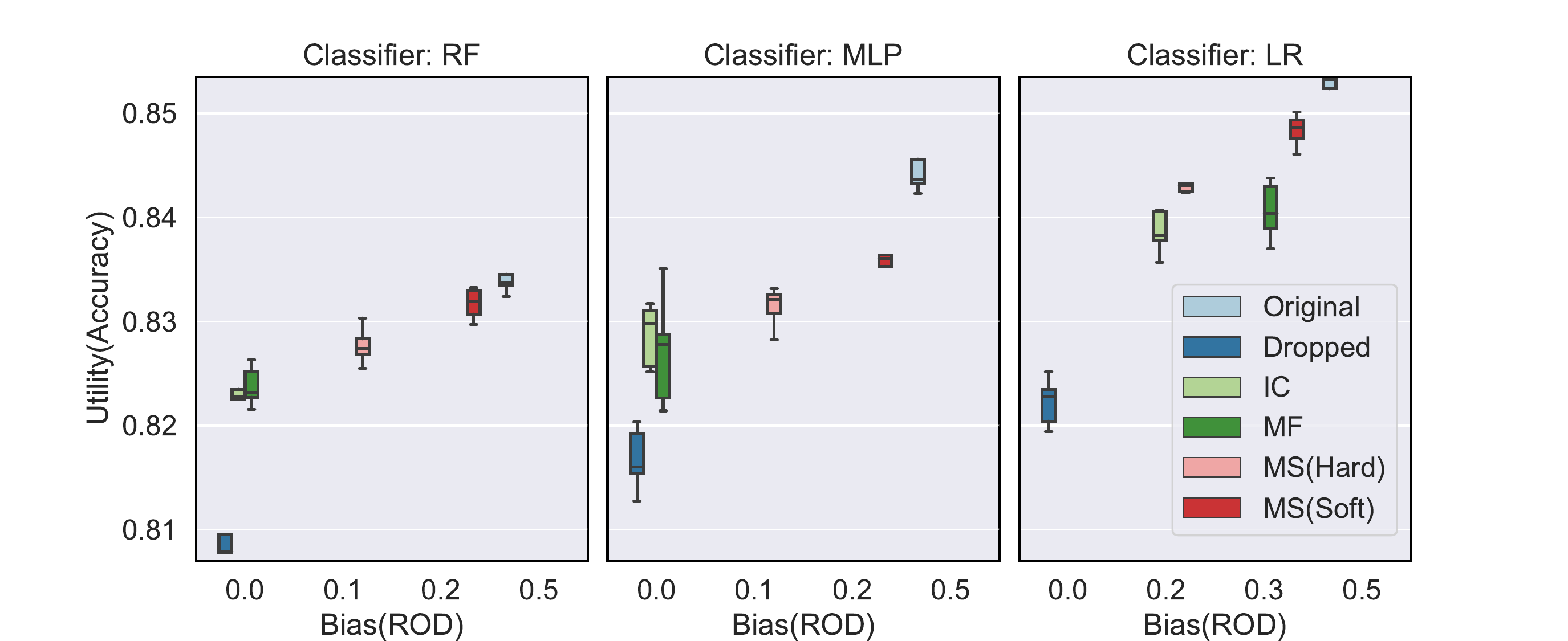}
	\caption{\textmd{Performance of \sys\ on Adult data.}}
				\label{fig:adult}
\end{figure}
\vspace{-0.3cm}
\begin{example} \label{ex:capuchin} Suppose financial organisations use the Adult data described in Ex~\ref{ex:berkeley} to train an ML model to assist them in verifying the reliability of their customers. The use of raw data for training an ML model leads to a model that is discriminative against females simply because the model picks up existing bias in data, as described in Ex~\ref{ex:adult}. To remove direct and indirect effects of gender on income (the red paths from $G$ to $Y$ in Fig.~\ref{fig:adult}(b)) using the \sys\ system, it is sufficient to enforce the CI $(O \indep \mb S, \mb M| \mb C,\mb E, \mb W)$ in data. Then, any model trained on the repaired data can be shown to be justifiably fair even on  unseen test data under some mild assumptions \cite{salimi2019interventional}. To empirically assess the efficacy of the \sys\ system, we repaired Adult data using the following \sys algorithms:  Matrix Factorization (MF), Independent Coupling (IC), and two versions of the MaxSAT approach: MS(Hard), which strictly enforces a CI, and MS(Soft), which approximately enforces a CI.  
Then, three classifiers -- Linear Regression (LR), Multi-layer Perceptron (MLP), and Random Forest (RF) -- were trained on both original and repaired training datasets using the set of variables $ \mb A \cup \mb N \cup \mb S$. The classifier also trained on raw data using only $\mb A$, i.e., we dropped the sensitive and inadmissible variables. The utility and bias metrics for each repair method were measured using five-fold cross validation. Utility was measured by the classifiers' accuracy, and bias measured by the Ratio of Observational discrimination introduced in \cite{salimi2019interventional}, which quantifies the effect of gender on outcome of the classifier by controlling for admissible variables (see \cite{salimi2019capuchin} for details).   Fig.~\ref{fig:adult} compares the  utility and bias of \sys\ repair methods on Adult data.  As shown, all repair methods successfully reduced
the ROD for all classifiers. The \sys\ repair methods had an effect similar to dropping the sensitive and inadmissible variables completely, but they delivered much higher accuracy (because the CI was enforced approximately).  
\end{example}
\vspace{-0.3cm}
\paragraph{Future Extensions.}
The problem of repairing data w.r.t a set of CI constraints was studied in \cite{salimi2019interventional} for a single saturated CI constraint problem.\footnote{A CI statement is saturated if it contains all attributes.}  In the presence of multiple training labels and sensitive attributes, one needs to enforce multiple potentially interacting or inconsistent CIs; this is more challenging and requires further investigation. In addition, further research is required on developing approximate repair methods to be able to trade the fairness and accuracy of different ML applications.


\vspace{-0.3cm}
\subsection{Fairness-Aware Weak Supervision Methods}
ML pipelines rely on massive labeled training sets. In most practical settings, such training datasets either do not exist or are very small. Constructing large labeled training datasets can be expensive, tedious, time-consuming or even impractical. This has motivated a line  of work on developing techniques for addressing the data labeling bottleneck, referred to as {\em weak supervision methods}. The core idea is to programmatically label training data using, e.g., domain heuristics \cite{ratner2016data}, crowdsourcing \cite{raykar2010learning} and distant supervision \cite{mintz2009distant}. In this context, the main challenges are handling noisy and unreliable sources that can potentially generate labels that are in conflict and highly correlated.  State-of-the-art frameworks for weak supervision, such as Snorkel \cite{ratner2017snorkel}, handle these challenges by training label models that take advantage of conflicts between all different labeling sources to estimate their accuracy. The final training labels are obtained by combining the result of different  labeling sources weighted by their estimated accuracy. While the focus of existing work is on collecting quality training labels to maximize the accuracy of ML models, the nuances of fairness cannot be captured by the 
exiting machinery to assess the reliability of the labeling sources. In particular, a new set of techniques is required to detect and explain whether certain labeling sources are biased and to combine their votes fairly.

\subsection{Provenance for Explanation}
\textit{Data provenance} refers to the origin, lineage, and source of data. Various data provenance techniques have been proposed to assist researchers in understanding the origins of data \cite{glavic2007data}. Recently, data provenance techniques has been used to explain why integrity constraints fail \cite{xu2018provenance}. These techniques are not immediately applicable to fairness integrity constraints, which are probabilistic. This motivates us to extend provenance to fairness or probabilistic integrity constraints in general. This extension is particularly 
crucial for reasoning about the fairness of training data collected from different sources by data integration and fusion, and it opens the opportunity to leverage existing techniques, such as provenance summarization
\cite{ainy2015approximated}, why-not provenance \cite{chapman2009not}, and query-answers causality and responsibility \cite{meliou2010complexity, DBLP:conf/icdt/SalimiB15,salimi2016quantifying,bertossi2017causes},  explanations for database queries queries \cite{roy2014formal}  to generate fine- and coarse-grained explanations for bias and discrimination.   
\vspace{-0.3cm}
\section{Conclusions}
This paper initiated a discussion on applying data management techniques in the embedding areas of algorithmic fairness in ML. We showed that fairness requires causal reasoning to capture natural situations, and that popular associational definitions in ML can produce incorrect or misleading results. 

\bibliographystyle{plain}
\small
\bibliography{ref}
\end{document}